\documentclass[fdp,fleqn]{w-art}
\usepackage{times}
\usepackage{w-thm}
\usepackage[]{graphicx}
\begin{document}
%%    The information for the title page will be placed between
%%    \begin{document} and \maketitle. The order of most entries
%%    is determined by the class file and can not be changed by
%%    rearranging them. The maketitle command follows after the
%%    abstract.
%%
%%    Most of the following commands will be completed by the publisher.
%%
%%    The copyrightyear is defined in the .clo file as the first argument
%%    of the copyrightinfo command. If the copyrightyear differs from that
%%    value it might be adjusted by the following definition:
%%
%% \renewcommand{\copyrightyear}{2003}% uncomment to change the copyrightyear.
%%
\DOIsuffix{theDOIsuffix}
%%
%% issueinfo for header and copyright line
\Volume{55}
\Issue{1}
\Month{01}
\Year{2007}
%%
%%    First and last pagenumber of the article. If the option
%%    'autolastpage' is set (default) the second argument may be left empty.
\pagespan{3}{}
%%
%%    Dates will be filled in by the publisher. The 'reviseddate' and
%%    'dateposted' (Published online) entry may be left empty.
%\Receiveddate{ November 2007}
%\Reviseddate{30 November 2007}
%\Accepteddate{2 December 2007}
%\Dateposted{3 December 2007}
%%
%\keywords{List, of, comma, separated, keywords.}

%% \pretitle{Editor's Choice}

%% We have a short and a long form for the title. The short form
%% (optional argument) goes into the running head.

\title[ F-$SU(5)$]{Yukawa couplings in F-theory $SU(5)$}

%% Please do not enter footnotes or \inst{}-notes into the optional
%% argument of the author command. The optional argument will go into
%% the header.  If there is only one address the marker \inst{x} may be
%% omitted.

%% Information for the first author.
\author[G.K. Leontaris:]{G.K. Leontaris\inst{1,}%
  \footnote{Corresponding author\quad E-mail:~\textsf{leonta@uoi.gr} ,
            Phone: +30\,2651\,008\,744,
            Fax: +30\,2651\,008\,698}}
\address[\inst{1}]{Theoretical Physics Division, Ioannina
University, GR-45110 Ioannina, Greece}
%%
%%    Information for the second author
%%  %%\author[S. Author]{Second Author\inst{1,2,}\footnote{Second author footnote.}}
%%  %%\address[\inst{2}]{Second address}
%%
%%    Information for the third author
%% %%\author[T. Author]{Third Author\inst{2,}\footnote{Third author footnote.}}
%%
%%    \dedicatory{This is a dedicatory.}
\begin{abstract}
\end{abstract}
%% maketitle must follow the abstract.
\maketitle                   % Produces the title.

%% If there is not enough space inside the running head
%% for all authors including the title you may provide
%% the leftmark in one of the following three forms:

%% \renewcommand{\leftmark}
%% {F. Author: A short title}

%% \renewcommand{\leftmark}
%% {F. Author and S. Author: A short title}

%% \renewcommand{\leftmark}
%% {F. Author et al.: A short title}

%% \tableofcontents  % Produces the table of contents.
\section{Abstract}

The fermion mass textures are discussed in the context of F-theory $SU(5)$ GUT.  The tree-level up, down and charged lepton Yukawa couplings are computed in terms of the integrals of overlapping wavefunctions  at the intersection points of three matter curves. All remaining entries in the fermion mass matrices can also be reliably estimated from higher order non-renormalizable Yukawa couplings  mediated by heavy string modes and/or Kaluza-Klein states.

\section{Wavefunction overlapping Integrals}

In F-theory GUTs  the trilinear Yukawa couplings are realised at the intersections of three matter curves  where the zero-modes of two fermion fields and a Higgs boson reside.  The structure of the zero-mode wavefunctions is found by solving the corresponding differential equations emerging from the twisted eight-dimensional Yang-Mills action~\cite{Beasley:2008dc}. The third generation  up, down and charged lepton mass matrices  are then computed in terms of the integrals of overlapping wavefunctions    at the intersection point of three matter curves~\cite{Cecotti:2009zf}. Furthermore, assuming that higher order non-renormalizable Yukawa couplings are generated through mediation of heavy string modes and/or Kaluza-Klein states, the calculation of  all entries in the fermion mass matrices can be reduced to a similar computation.  In fact,  in this approach a rigorous and consistent application of the described method can be developed~\cite{Leontaris:2010zd}, by computing the relevant integrals for the NR-terms involving zero-mode and massive KK-mode overlapping wavefuctions. The complete   calculation of all $3\times 3$ Yukawa entries for the entire charged fermion mass spectrum shows that  all the Yukawa coefficients are found within a reasonable range of values ($\le {\cal O}(1)$) while the
 predicted fermion mass eigenvalues and Cabibbo Kobayashi Maskawa (CKM)-mixing in accordance to the experimental expectations.
The tree-level Yukawa couplings are computed by the wavefunction overlap integrals
\begin{eqnarray}
\lambda_{ij}&=&\int_S\psi_i\,\psi_j\,\phi \;dz_1\wedge\,d\bar z_1\wedge dz_2\wedge\,d\bar z_2
\label{Yuk}
\end{eqnarray}
where $S$ is the compact 4-d manifold wrapped by 7-brane surface associated to the gauge symmetry (here $SU(5)$).  To compute the integral we use our knowledge of the wavefunctions' profile close to the intersection point.  For localized solutions  the zero-mode equations lead to a Gaussian profile
 of the general form
 \begin{eqnarray}
\psi
      &\propto&e^{-m^2\frac{|q_1z_1+q_2z_2|^2}{q}}\label{psi}
\end{eqnarray}
where  $m$ is a mass parameter related to a background Higgs vacuum expectation value (vev) $\langle \Phi\rangle$, $z_{1,2}$ complex coordinates on $S$ and $q_i$ are $U(1)$ charges ($q=\sqrt{q_1^2+q_2^2}$).
To make a reliable computation of the Yukawa couplings we also need a consistent  normalization of the wavefunction. For a generic form of a wavefunction  we have
\begin{equation}
{\cal C}=M_*^4\int_S\,|\psi|^2dz\wedge d\bar z\, =\,\pi \frac{M_*^4}{m^2q}{\cal R}^2\label{Cnorm}
\end{equation}
 The factor $\pi \frac{M_*^4}{m^2q}$ is the result of the gaussian integration along the coordinate
  normal to the curve. The factor ${\cal R}^2$ is introduced to account for the integration along the coordinate parametrising the curve.
The mass scale $m$ is naturally associated to the compactification scale  $m\sim M_*\approx M_C$ on $S$,
while the parameter ${\cal R}$ `measures' the integration along the  matter curve  inside
the GUT surface $S$, thus clearly ${\cal R}^{-1}<M_C$  and one naturally  expects that ${\cal R}^{-1}\approx M_{GUT}$. From  renormalization group analysis we can see~\cite{Leontaris:2011pu}
that it is possible to obtain a scale independent normalization given by
\[
\frac{1}{\sqrt{C}}=\sqrt{\frac{q}{\pi}}\;e^{2/3({\cal T}_{5/6}-{\cal T}_0)}
\]
where the exponent  ${\cal T}_{5/6}-{\cal T}_0$ is related to KK-massive mode threshold
corrections expressed in terms of the analytic torsion ${\cal T}$~\cite{Ray:1973sb}.
The  integral for three generic  wavefunctions  of the form (\ref{Yuk})
with arbitrary charges $q_i,q_i',q''_i, i=1,2$ participating in the triple intersection the
Yukawa couplings, gives~\cite{Leontaris:2010zd}
\begin{equation}
\lambda=e^{2({\cal T}_{5/6}-{\cal T}_0)}\frac{4\,\sqrt{\pi} }{q+q'+q''}\,\frac{(qq'q'')^{3/2}}{(q_1q_2'-q_1'q_2)^2}
\label{Yukint}
\end{equation}
where $q_i''=-q_i-q'_i$ from charge conservation in the triple intersection, while $q=\sqrt{q_1^2+q_2^2}$
etc.
\subsection{ Calculation of the bottom Yukawa coupling}

In the present context we have assumed that our local F-theory GUT is  $G_S=SU(5)$
with the top and bottom Yukawa couplings appearing in triple intersections of $S$ with other
seven branes. Locally the $G_S$ symmetry is enhanced and in practice the Yukawa
couplings should also be invariant  under additional $U(1)$ symmetries.
In the spectral cover approach we consider the $SU(5)$ embedding  into the exceptional symmetry
 $E_8\supset SU(5)_{GUT}\times SU(5)_{\perp}$ which is the highest symmetry in the elliptic fibration. We assume  the background field configuration  $\Phi$ with weights $t_i$ along $SU(5)_{\perp}$ satisfying the tracelessness condition $\sum_{i=1}^5t_i=0$.
We turn on a non-zero vev for $\Phi$ which breaks the symmetry  down to $SU(5)_{GUT}\times U(1)^{4}$.
 At the intersection points where the Yukawa couplings are formed two appropriate linear combinations
 of these $U(1)$'s are preserved.  The  $SU(5)$ representations are characterised by various
 combinations of $t_i$'s as shown in Table~\ref{rest}. An $SU(5)$ gauge invariant coupling must
 also comply with the cancellation of the $t_i$ charges. In practice this means that for a given
 allowed coupling either the  $t_i$ must come in pairs with opposite sign, or their sum  has to satisfy the traceless condition.
\begin{table}[tbp] \centering%
\begin{tabular}{|l|c|c|c|c|}
\hline
Field&$U(1)_i$& homology& $U(1)_Y$-flux&$U(1)$-flux\\
\hline
$10^{(1)}$& $t_{1,2}$& $\eta-2c_1-{\chi}$&$ -N$ &$M_{10_1}$\\ \hline
$10^{(2)}$& $t_{3}$& $-c_1+\chi_7$&$ N_7$ &$M_{10_2}$\\ \hline
$10^{(3)}$& $t_{4}$& $-c_1+\chi_8$&$ N_8$ &$M_{10_3}$\\ \hline
$10^{(4)}$& $t_{5}$& $-c_1+\chi_9$&$ N_9$ &$M_{10_4}$\\ \hline
$5^{(0)}$& $-t_{1}-t_2$& $-c_1+{\chi}$&$ N$ &$M_{5_{h_u}}$\\ \hline
$5^{(1)}$& $-t_{1,2}-t_3$& $\eta -2c_1-{\chi}$&$ -N$ &$M_{5_1}$\\ \hline
$5^{(2)}$& $-t_{1,2}-t_4$& $\eta -2c_1-{\chi}$&$ -N$ &$M_{5_2}$\\ \hline
$5^{(3)}$& $-t_{1,2}-t_5$& $\eta -2c_1-{\chi}$&$ -N$ &$M_{5_3}$\\ \hline
$5^{(4)}$& $-t_{3}-t_4$& $-c_1+{\chi}-\chi_9$&$N-N_9$ &$M_{5_4}$\\ \hline
$5^{(5)}$& $-t_{3}-t_5$& $-c_1+{\chi}-\chi_8$&$ N-N_8$ &$M_{5_{h_d}}$\\ \hline
$5^{(6)}$& $-t_{4}-t_5$& $-c_1+{\chi}-\chi_7$&$ N-N_7$ &$M_{5_6}$\\ \hline
\end{tabular}%
\caption{Matter curves available to accommodate the field representation content
under $SU(5)\times U(1)_{t_i}$, their homology class and flux restrictions~\cite{Dudas:2010zb,Leontaris:2010zd}.
A $Z_2$ monodromy $t_1\leftrightarrow t_2$ is imposed.}
\label{rest}
\end{table}
The aim now is to calculate the top and bottom Yukawa couplings in a specific F-theory construction.
Here the computation for the model~\cite{Dudas:2010zb,Leontaris:2010zd} where
both up and down quark mass matrices are rank one  in agreement with the hierarchical mass matrix structure is presented. In particular, we discuss the details for the bottom quark Yukawa coupling. In the model under consideration this is  obtained  from a trilinear term when families are accommodated according to
 ${ 10}^{(1)}\rightarrow {\bf 10}_3$, $\bar 5^{(5)}\rightarrow {\bf \bar 5}_{3}$,  $\bar 5^{(2)}\rightarrow {\bf \bar 5}_{h_d}$, and a $Z_2$ monodromy $t_1\leftrightarrow t_2$ is assumed, so that
 \begin{eqnarray}
W_{tree}&=&{\bf 10}_3\;\;\;\cdot {\bf \bar 5}_{3}\;\;\;\cdot\;\;\; {\bf \bar 5}_{h_d}
\nonumber\\
&&            \;t_2,\;\;\;\; t_3+t_5,\;t_1+t_4\nonumber
\end{eqnarray}
 where in the second line the $t_i$-combinations for the corresponding
 $SU(5)_{GUT}$ representations are shown. Clearly this $SU(5)$ invariant coupling satisfies
 the condition $\sum_it_i=0$ and therefore is allowed at tree-level.
  Recall now that the  generators of the $U(1)$s are given by
 the  four diagonal generators, $Q_i$,  of $SU(5)_{\perp}$. It is convenient to introduce the basis
 column vectors ${|t_i>}_j=\delta_{ij}$, $i,j=1,\dots 5$.
 % or explicitly
%\begin{eqnarray}
%\begin{align}
%|t_1>=\left(\begin{array}{c}1\\0\\0\\0\\0\end{array}\right),\;
%|t_2>=\left(\begin{array}{c}0\\1\\0\\0\\0\end{array}\right),\;
%|t_3>=\left(\begin{array}{c}0\\0\\1\\0\\0\end{array}\right),\;
%|t_4>=\left(\begin{array}{c}0\\0\\0\\1\\0\end{array}\right),\;
%|t_5>=\left(\begin{array}{c}0\\0\\0\\0\\1\end{array}\right),\;
%\end{align}
%\end{eqnarray}
The local form of  $Q_i$'s  is determined by demanding that the two of them annihilate the states participating
in the vertex where the bottom Yukawa coupling is realized. In other words,  for a given Yukawa coupling the fields involved will have zero eigenvalue for two combinations of the charge generators. Once we have identified them we  easily construct  the two orthogonal to them operators. Then the normalized charges of the fields are determined by the action of these operators on the corresponding states.

Recall now  that the interacting fields for the bottom Yukawa have weights $t_{1,2},\;t_{2,1}+t_4$ and $t_3+t_5$. We first observe that the two operators
\begin{eqnarray}
Q_3=\frac{1}{\sqrt{2}}\{0,0,1,0,-1\},\;\; Q_4=\frac{1}{\sqrt{2}}\{0,1, 0,-1,0\}
\end{eqnarray}
 annihilate  the  states participating in the bottom quark vertex
\[
Q_{3,4}|t_{1,2}>=0,\; Q_{3,4}|t_{1,2}+t_4>=0,\; Q_{3,4}|t_3+t_5>=0\ \cdot
\]
Thus, $Q_1,Q_2$ operators are part of a suitable $4d$ basis for vertices involving the above $t_i$-charge combinations.
The most general vectors which fulfill $\sum t_i=0$ and at the same time  are orthogonal to them are
\begin{eqnarray}
Q_1&=&\frac{1}{\sqrt{2}{\cal D}}\{-2 (a+b),a,b,a,b\}\label{vec3}\\
Q_2&=&\frac{1}{\sqrt{10}{\cal D}}\{2 (a-b),2 a+3 b, -3 a-2 b, 2a+3 b,-3 a-2 b\}\label{vec4}
\end{eqnarray}
with  ${\cal D}^2=3a^2+4ab+3b^2$. For example, choosing $a=-b=1$,  then $
Q_3=\frac{1}{2}\{0,1,-1,1,-1\}$ and
$Q_4=\frac{1}{2\sqrt{5}}\{4, -1,-1,-1,-1\}$
 corresponding to the $U(1)$ charge assignment obtained under the chain
\[SU(5)_{\perp}\rightarrow SU(4)\times SU(1)\rightarrow\cdots\rightarrow U(1)^4\]
Choosing $a=0,b=1$, we get $Q_3=\frac{1}{2}\{-2,0,1,0,1\},\;
Q_4=\frac{1}{\sqrt{30}}\{-2,3,-2,3,-2\}$
 corresponding to another linear combination of the $U(1)$ charge assignment obtained under the
 breaking pattern $SU(5){\perp}\rightarrow SU(3)\times SU(2)\rightarrow \cdots\rightarrow U(1)^4$.
Acting on the $|t_1>$ and $|t_2+t_4>$ states with these operators we obtain the charges
\begin{eqnarray}
\{q_1,q_2\}=\left\{
-2\frac{a+ b}{\sqrt{2}\,{\cal D}},2\frac{ a-b}{\sqrt{10} \,{\cal D}}\right\},\;\;
\{q'_1,q'_2\}=\left\{
2\frac{a}{\sqrt{2}\,{\cal D}},2\frac{2a+3b}{\sqrt{10} \,{\cal D}}\right\}
\end{eqnarray}
Then, the quantities appearing in (\ref{Yukint}) are
\begin{eqnarray}
q=\sqrt{q_1^2+q_2^2}=\frac{2}{\sqrt{5}},\,\;
q'=\sqrt{{q'_1}^2+{q'_2}^2}=\sqrt{\frac{6}{5}},\,\;
q_1'q_2-q_1q_2'=-\frac{2}{\sqrt{5}}\nonumber
\end{eqnarray}
while $q'' $ can be written in terms of $q_i,q''_i$ by charge conservation.
We observe that the parameters involved in the computed integral are independent of $a,b$, and
the specific breaking pattern of $SU(5)_{\perp}$.
To compute the value of the overlapping wavefunctions integral and demonstrate the points discussed
above, we proceed with the computation  of the torsion ${\cal T}$  in a simple case.
As in ref~\cite{Donagi:2008kj} we take a line bundle ${\cal O}(n,-n)$ on a Hirzebruch surface
$F_0=P^1\times P^1$. After some algebra we get~\cite{Leontaris:2010zd}
\[{\cal T}_{5/6}-{\cal T}_0=\zeta'_0(0)-\zeta'_{-1}(0)=-\frac 12\]
The resulting bottom quark coupling is  $\lambda_b\approx 1.17$  while repeating the above
 analysis one finds  $\lambda_t\approx 1.23.$ Thus, in this simple example, the values
 of the third generation Yukawa couplings result to  $m_t,m_b$ masses which are close
 to the experimental findings.
\begin{table}[!h]
\centering
\begin{tabular}{|l|r|r|rrr||l|r|r|rr|}
\hline
\multicolumn{11}{|c|}{Chiral Matter}\\
\hline
\hline
     &  $M$  &  $N$  &  $Q$  &  $u^c$  &  $e^c$ &     &  $M$  &  $N$  &  $d^c$&  $L$    \\
\hline
$10^{(1)}\,(F_3)$ &  1  &  0  &  1  &  1  &  1        &$5^{(4)}\,({\bar f}_1)$  & -1  &  0  &  -1 & -1 \\
$10^{(2)}\,(F_{2,1})$ &  1  &  -1 &  1  &  2  &  0  &$5^{(1)}\,({\bar f}_2)$  &  -1 &  0  & -1  & -1   \\
$10^{(3)}\,(F_{1,2})$ &  1  &  1  &  1  &  0  &  2  &$5^{(2)}\,({\bar f}_3)$  &  -1 &  0  & -1  & -1 \\
$10^{(4)}\,(-)$ &  0  &  0  &  0  &  0  &  0         &$5^{(3)}\,(-)$  & 0   &  0  &  0  &  0   \\
\hline
\end{tabular}
\begin{tabular}{lrrrr}
     &  &    &  &
\end{tabular}
\begin{tabular}{|l|r|r|rr|}
\hline
\multicolumn{5}{|c|}{Higgs and Colour Triplets}\\
\hline
\hline
     &  $M$  &  $N$  &  $T$&  $h_{u,d}$    \\
\hline
$5^{(0)}\,(h_u,T)$&  1    &  0    &  1    &  1\\
\hline
$5^{(5)}\,(h_d)$  & 0   &  -1 &  0  & -1  \\
\hline
$5^{(6)}\,(\bar T)$  & -1  &  1  & -1  & 0\\
\hline
\end{tabular}
\caption{The distribution of the chiral and Higgs matter content of the minimal
model along the available curves, after the $U(1)_Y$ flux is turned on. }
\label{content}
\end{table}
The  remaining  mass entries for the up and down are generated from non-renormalizable contributions.
They emerge through the mediation of KK-massive modes and all Yukawa coefficients
are calculable using the above procedure.  Here we give a specific model with
representation content given in Table \ref{content} which is also  capable of generating doublet-triplet splitting through the flux mechanism. Thus the up and down quark  mass matrices
have the form
\[
M_u  = \left( {\begin{array}{*{20}c}
   { \theta _{14} ^2 \theta _{43}^2 } & {\theta _{14} ^2 \theta _{43}^{} } & { \theta _{14} ^{} \theta _{43}^{} }  \\
   {\theta _{14} ^2 \theta _{43}^2 } & {\theta _{14} ^2 \theta _{43}^{} } & {\theta _{14} ^{} \theta _{43}^{} }  \\
   {\theta _{14} ^{} \theta _{43}^{} } & {\theta _{14} } &\sim 1  \\
\end{array}} \right),\;
M_d=\left(
\begin{array}{lll}
\theta _{14}^2 \theta _{43}^2 & \theta _{14} \theta _{43}^2 &  \theta _{14} \theta _{43} \\
 \theta _{14}^2 \theta _{43} & \theta _{14} \theta _{43} &\,\theta _{14} \\
 \theta _{14} \theta _{43} &\theta _{43} &\sim 1
\end{array}
\right)
\]
where  $\theta_{ij}$ represent  $SU(5)$ singlet field vevs. It can be
seen~\cite{Leontaris:2010zd} that the above are in accordance with the quark
mass ratios and CKM mixing. Moreover, a similar structure is obtained for the charged
lepton mass matrix, whilst the neutrinos can be made consistent with tri-bi maximal mixing.

\end{document}